\pdfoutput=1
\documentclass{emulateapj}
\usepackage{color}

\newcommand{\etal}{{et al.\/}}

\shorttitle{$\gamma$-ray Variability In Blazars}
\shortauthors{MacDonald \etal}

\begin{document}

\title{Through the Ring of Fire: $\gamma$-ray Variability In Blazars By A Moving Plasmoid Passing A Local Source Of Seed Photons}

\author{Nicholas R. MacDonald\altaffilmark{1}, Alan P. Marscher\altaffilmark{1}, Svetlana G. Jorstad\altaffilmark{1,2} and Manasvita Joshi\altaffilmark{1}}
\affil{$~^{1}$Institute for Astrophysical Research, Boston University, 725 Commonwealth Avenue, Boston, MA 02215; nmacdona@bu.edu}
\affil{$~^{2}$Astronomical Institute, St. Petersburg State University, Universitetskij Pr. 28, Petrodvorets, 198504 St. Petersburg, Russia}

\begin{abstract}
Blazars exhibit flares across the electromagnetic spectrum.  Many $\gamma$-ray flares are highly correlated with flares detected at optical wavelengths; however, a small subset appears to occur in isolation, with little or no variability detected at longer wavelengths.  These ``orphan'' $\gamma$-ray flares challenge current models of blazar variability, most of which are unable to reproduce this type of behavior.  We present numerical calculations of the time-variable emission of a blazar based on a proposal by \citeauthor{marscher10} to explain such events.  In this model, a plasmoid (``blob'') propagates relativistically along the spine of a blazar jet and passes through a synchrotron-emitting ring of electrons representing a shocked portion of the jet sheath.  This ring supplies a source of seed photons that are inverse-Compton scattered by the electrons in the moving blob.  The model includes the effects of radiative cooling, a spatially varying magnetic field, and acceleration of the blob's bulk velocity.  Synthetic light curves produced by our model are compared to the observed light curves from an orphan flare that was coincident with the passage of a superluminal knot through the inner jet of the blazar PKS 1510$-$089.  In addition, we present Very Long Baseline Array polarimetric observations that point to the existence of a jet sheath in PKS 1510$-$089, thus providing further observational support for the plausibility of our model.  An estimate of the bolometric luminosity of the sheath within PKS 1510$-$089 is made, yielding $L_{\rm sh} \sim 3 \times 10^{45} ~ \rm erg ~ \rm s^{-1}$.  This indicates that the sheath within PKS 1510$-$089 is potentially a very important source of seed photons.
\end{abstract}


\section{Introduction}

Blazars are the most luminous persistent objects in the sky (up to $\sim 10^{48}$ erg  s$^{-1}$; e.g., \citealt{krolik99}).  They exhibit
variability across the electromagnetic spectrum on timescales ranging from minutes to years  (e.g., \citealt{aharonian07} and references therein).  Blazars constitute a sub-class of active galactic nuclei (AGNs) whose relativistic plasma jets are thought to be closely aligned 
to our line of sight \citep{blandford79}.  The emission from blazars is predominantly 
non-thermal, including radio synchrotron radiation and X-ray and $\gamma$-ray inverse-Compton scattering.  
The central engines of blazars, which are believed to be the ultimate source of this variability, are unresolved and opaque to radio waves.  Hence, they are not imaged directly, even with the $0.1$ mas resolution obtained with the Very Long Baseline Array (VLBA) at 43 GHz.  Despite this limitation, by monitoring and subsequently modeling flares in the high-energy emission from these AGN, 
we can potentially gain insight into the sub-parsec-scale physics of the jets close to the central engines.

Many theoretical models have been developed to explain both the spectral energy distributions (SEDs) and high-energy variability of blazars.  The SEDs of most blazars take the form of two humps - one in the radio/UV portion of the spectrum (synchrotron) and the other at higher energies in the X-ray/$\gamma$-ray portion of the spectrum (interpreted here as inverse-Compton emission).  \citet{blandford79}, \citet{marscher85}, and \citet{hughes85} proposed that shock acceleration within the jet could result in the observed variations in radio emission.  \citet{dermer93} presented a leptonic model in which the high-energy hump of the SED is produced by comptonization of accretion disk photons by electrons contained within a plasmoid (``blob'') moving relativistically along the jet.  Alternatively, \citet{bottcher09} formulated a hadronic model in which the high-energy hump is produced through $p\gamma$ pion production from relativistic protons within the jet encountering both internal and external photon fields.  \citet{sikora94} also explored the effects of comptonization of external radiation fields from the accretion disk and the surrounding broad emission-line region (BLR) by electrons within the jet.  \citet{spada01} presented an internal shock model for blazar variability in which shells of plasma within the jet collide, producing forward and reverse shocks that result in particle acceleration within the jet.  While these models can explain many aspects of the radiative mechanisms potentially at work in blazars, there remain aspects of high-energy variability that are not accounted for by these theories.

\begin{figure}
\epsscale{1.15}
\plotone{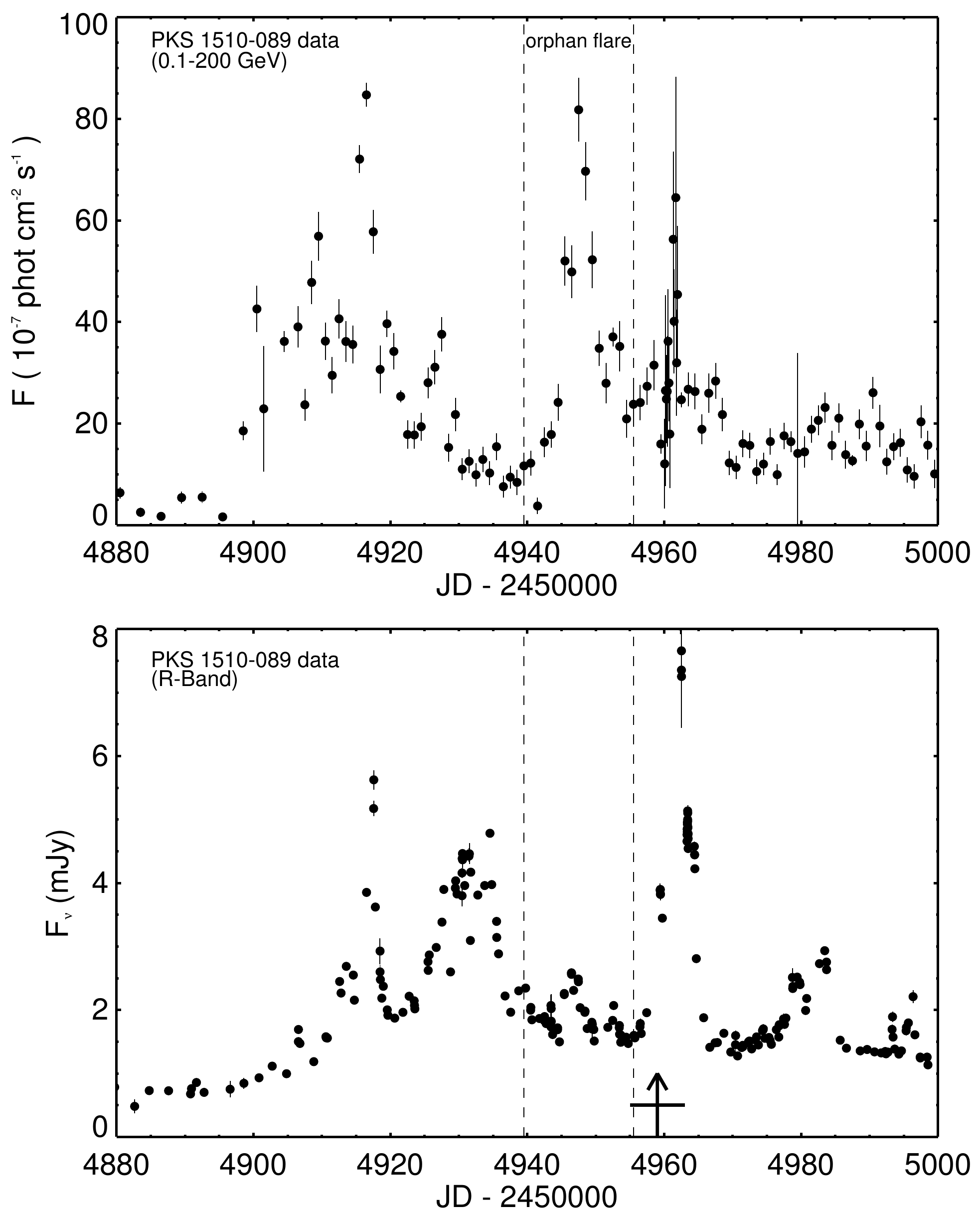}
\caption{\label{fig1}$\gamma$-ray (upper panel) and optical (lower panel) light curves of PKS 1510$-$089 during 2009.  The dashed vertical lines delineate an example of an orphan $\gamma$-ray flare in which the prominent flare in the GeV band has at most a weak counterpart in the optical band.  The vertical arrow in the lower panel marks the time when a superluminal knot passed through the 43 GHz core of PKS 1510$-$089, coincident with a large optical/$\gamma$-ray flare that occurred roughly 20 days after the onset of the orphan flare \citep{marscher10}.  The horizontal bar represents the uncertainty in the time of the knot's passage through the radio core.}
\end{figure}

Recent observations obtained with the \textit{Fermi} Large Area Telescope (LAT) and the VLBA have shown that a large number of $\gamma$-ray flares in blazars are strongly correlated with the passage of superluminal knots through the radio cores of these objects \citep{marscher12}.  The majority of these $\gamma$-ray flares tend to be associated with flares detected at optical/UV wavelengths.  The high cadence of observations provided by the \textit{Fermi} spacecraft, however, has also allowed the identification of a small sub-set of $\gamma$-flares that seem to occur in relative isolation, with little or no variability detected in the other bands.  These isolated events are termed ``orphan'' flares.  In one case, \citet{marscher10} present observations of an orphan $\gamma$-ray flare that occurred during the passage of a superluminal knot through the inner jet of the blazar PKS 1510$-$089 (see Figure \ref{fig1}).

Based on the orphan $\gamma$-ray flare highlighted in Figure \ref{fig1}, here we develop a model of blazar variability to account for this type of behavior.  In our model, a blob consisting of a power-law distribution of electrons propagates relativistically along the jet axis of the blazar and passes through a 
synchrotron-emitting ring of electrons, which represents a shocked segment of a more slowly moving jet sheath (see Figure \ref{fig2} and the sketch presented in Figure \ref{fig9} below).  The ring creates a very localized source of seed photons that are inverse-Compton scattered by the electrons in the moving blob, creating an orphan $\gamma$-ray flare as the blob passes through the ring.  We point out that our model is distinct from Compton mirror models (see e.g., \citealt{ghisellini96}, \citealt{bottcher98}, and \citealt{bottcher05}) that require a reflecting cloud close to the jet axis to produce $\gamma$-ray flares. 

The presence of a non- to mildly-relativistic sheath of plasma enshrouding the relativistic spine of a blazar jet has been discussed by several authors.  From a theoretical standpoint, \citet{ghisellini05} investigated the radiative interplay between these two regions within the jet via the inverse-Compton mechanism.  More recently, the SEDs of several blazars have been fit with models that utilize the presence of a jet sheath (see, e.g., \citealt{alesic14}  and \citealt{tavecchioa14}).  In addition, \citet{tavecchiob14} discuss the potential of sheaths of plasma within blazars to produce the high-energy neutrinos being detected by IceCube. Several authors have also discussed what the observational signature of a sheath of plasma would be within a blazar jet \citep[see][]{attridge99}.  Velocity shear between the jet and the ambient medium into which the jet propagates is expected to align the magnetic field on the outer edges of the jet with the jet axis so that the sheath polarization should be high (tens of percent) relative to the spine.  The electric vector position angles (EVPAs) associated with this polarization should be perpendicular to the jet axis.  This polarimetric signature has indeed been detected in a number of blazars \citep[see][]{pushkarev05}.  We present polarimetric observations of PKS 1510$-$089 that point to the existence of a jet sheath within this blazar, lending support to our model of orphan $\gamma$-ray flares.      

\begin{figure}
\epsscale{1.15}
\plotone{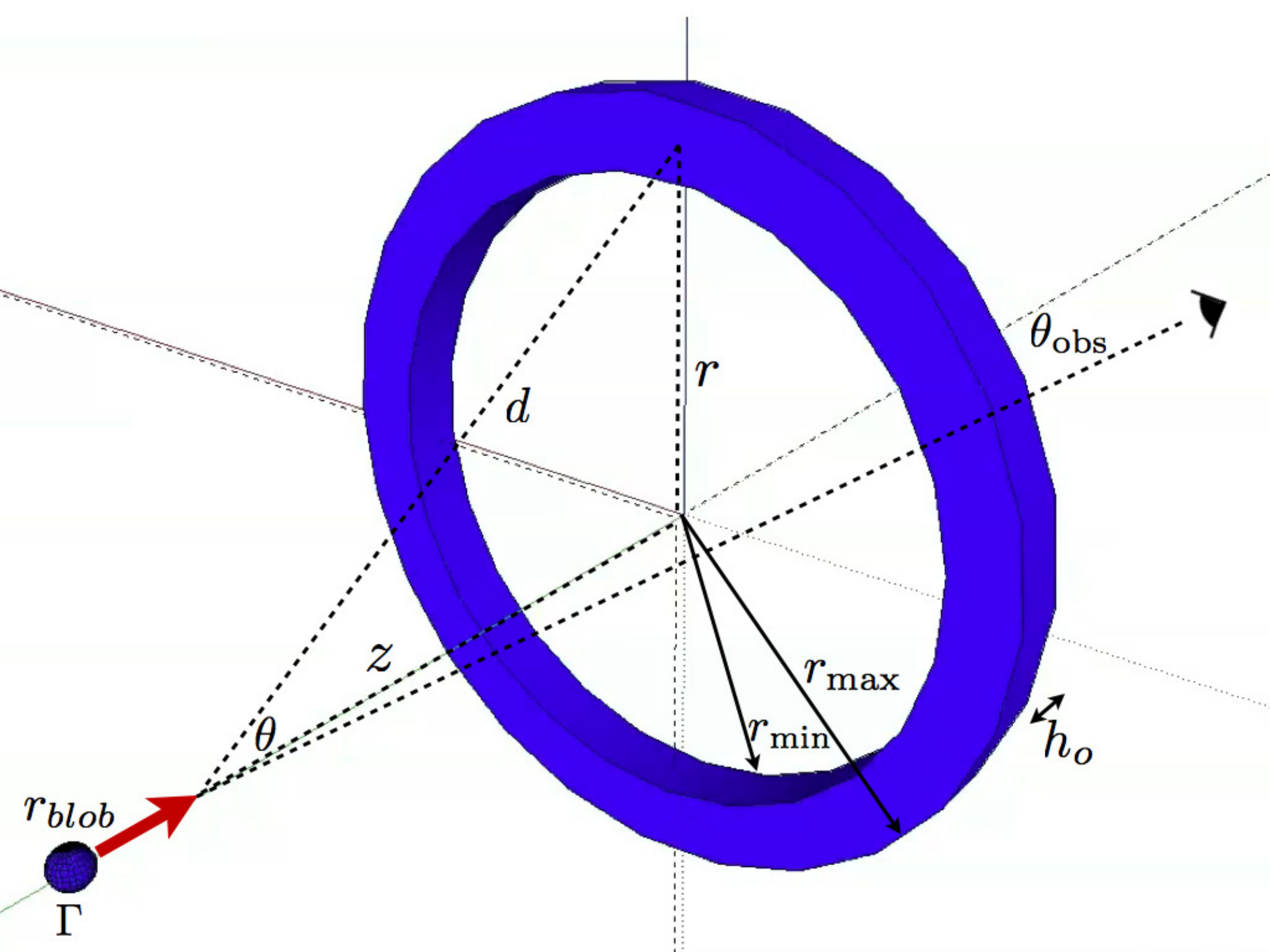}
\caption{\label{fig2}Proposed model for an orphan $\gamma$-ray flare: a blob containing relativistic electrons with a power-law distribution of energies passes through a synchrotron emitting ring and inverse-Compton scatters the ring photons.}
\end{figure} 

This paper is organized as follows: In \S2 we outline the theory relating to the radiative transfer modeled in our code.  In \S3 we describe the numerical simulation performed and list all of our model parameters.  In \S4 we compare synthetic light curves produced by the model to the observed optical and $\gamma$-ray light curves of the orphan flare of PKS 1510$-$089.  In \S5 we present VLBA polarimetric observations that point to the existence of a jet sheath within PKS 1510$-$089.  In \S6 we present our conclusions and summary.  We adopt the following cosmological parameters: $H_{\rm o} = 70 ~ \rm km ~ \rm s^{-1} ~ \rm Mpc^{-1}$, $\Omega_{\rm m} = 0.3$, and $\Omega_{\Lambda} = 0.7$.

\section{Theory}

Models of blazar emission, whether leptonic or hadronic, rely on the presence of an emission region (a ``blob'') moving relativistically along the jet.  For the purposes of this work we adopt the leptonic scenario and treat the blob (and ring) as consisting of electrons with a power-law energy distribution, $n_{e}(\gamma) \propto \gamma^{-s}$, over the range  $\gamma_{\rm min}$ to $\gamma_{\rm max}$, where the energy of a given electron is $E = \gamma m_{e} c^{2}$.

There are three main components of the non-thermal emission produced by the passage of a blob of plasma down the spine of a jet in our model.
The first is synchrotron radiation (syn) produced by electrons gyrating around magnetic field lines within the blob.  
These electrons also inverse-Compton scatter the synchrotron photons they produce, 
a process known as synchrotron self-Compton (ssc) emission.  The third component is
inverse-Compton scattering of the external (to the blob) ring synchrotron photons by the blob's electrons, 
a process known as external Compton scattering (ec).  As the blob approaches the ring, the photon density in the co-moving frame of the plasma increases, resulting in an ec driven $\gamma$-ray flare that peaks and decays as the blob passes through and then moves away from the ring. 

The photon production rates $(\rm photons ~ s^{-1} ~ cm^{-3} ~ \epsilon^{-1})$ for these three physical processes, $\dot{n}_{\rm syn}$, $\dot{n}_{\rm ssc}$, and $\dot{n}_{\rm ec}$, are given below ($\epsilon \equiv  h \nu / m_{\rm e} c^{2} $ is defined as a dimensionless photon energy).  In our radiative transfer code, we evaluate these photon production rates in the co-moving frame of the blob, after which a series of Lorentz transformations is applied to obtain the resultant flux in the observer's frame.  In all of the following equations, the dimensionless electron velocity, $\beta_{\rm e} \equiv v_{\rm e}/c  \approx 1$, has been set to unity.

The synchrotron photon production rate per unit volume for electrons within the blob or the ring is given by:
\begin{equation} \label{eq1}
\dot{n}_{\rm syn}(\epsilon) = \frac{\sqrt{3}e^{3}B}{2\pi h m_{e} c^{2} } \left( \frac{1}{ \epsilon } \right) \int_{\gamma_{\rm min}}^{\gamma_{\rm max}} \! F(x) ~ n_{\rm e}(\gamma) ~ \, \mathrm{d} \gamma
\end{equation}
(e.g., \citealt{joshi11}), where $x \equiv \nu/\nu_{\rm c}$ is the ratio of the observed frequency to the critical
frequency ($\nu_{\rm c}$), which is given by $\nu_{\rm c } \equiv  3 e B \gamma^{2} / 4 \pi m_{e} c ~$; and $F(x)$ is an approximation to a Bessel function \citep[see][]{joshi11}.

The ssc photon production rate per unit volume within the blob is given by:
\begin{equation} \label{eq2}
\dot{n}_{\rm ssc}(\epsilon) = \frac{1}{4\pi} \int_{\gamma_{\rm min}}^{\gamma_{\rm max}} \! n_{\rm e}(\gamma) \, \mathrm{d} \gamma \\
\int_{\epsilon_{\rm min}}^{\epsilon_{\rm max}} \! n_{\rm ph}^{\rm blob}(\epsilon) ~ g(\epsilon_{s},\epsilon,\gamma) ~ \, \mathrm{d} \epsilon
\end{equation}
(e.g., \citealt{joshi11}), where $n_{\rm ph}^{\rm blob}(\epsilon) = \dot{n}_{\rm syn}(\epsilon) ~ r_{blob} / c $ is the synchrotron photon
density produced by the electrons in the co-moving frame of the blob.  The isotropic scattering cross-section, $g(\epsilon_{s},\epsilon,\gamma)$, is given by:
\begin{eqnarray}\nonumber
g(\epsilon_{\rm s}, \epsilon, \gamma) = \left\{ \begin{array}{ll}
   \frac{3c \sigma_{\rm T}}{16 \gamma^{4} \epsilon} \left(\frac{4 
\gamma^{2} \epsilon_{\rm s}}{\epsilon} - 1\right)  & \textrm{for 
$\frac{\epsilon}{4 \gamma^{2}} \leq \epsilon_{\rm s} \leq \epsilon$}\\
\\ 
   \frac{3c \sigma_{\rm T}}{4 \gamma^{2} \epsilon} \left[\zeta_{1} + 
\zeta_{2} +\zeta_{3}\right]  ~ & \textrm{for 
$\epsilon \leq \epsilon_{\rm s} \leq \frac{4 \epsilon \gamma^{2}}{1 + 
4 \epsilon \gamma}$},
\end{array} \right.
\end{eqnarray}
 where $\sigma_{\rm T}$ is the Thomson cross-section, $\epsilon$ is the incoming photon energy, $\epsilon_{\rm s}$ is the scattered photon energy, and $\zeta_{1}$, $\zeta_{2}$, and $\zeta_3$ are defined as follows:
\begin{eqnarray}\nonumber \label{eq3}
\zeta_{1} &\equiv& 2\eta ~ \rm ln(\eta) \\
\zeta_{2} &\equiv& (1+2\eta)(1-\eta) \nonumber \\
\zeta_{3} &\equiv& \frac{ \left(4 \epsilon \gamma \eta\right)^{2} (1 - \eta)}
{2 \left(1 + 4 \epsilon \gamma \eta\right)} ~,
\end{eqnarray}
where $\eta \equiv \epsilon_{\rm s} / 4 \epsilon \gamma^{2} (1-\epsilon_{s}/\gamma)$ \citep{joshi11}.

The general ec photon production rate per unit volume within the blob for the scenario presented in Figure \ref{fig2} is given by:
\begin{eqnarray}\nonumber \label{eq4}
\dot{n}_{\rm ec}(\epsilon) &=& c
\oint \! \mathrm{d} \Omega
\int_{\epsilon_{\rm min}}^{\epsilon_{\rm max}} \! n_{\rm ph}^{\rm ring}(\epsilon) ~ \mathrm{d} \epsilon  
\oint \! \mathrm{d} \Omega_{\rm{e}} 
\int_{\gamma_{\rm min}}^{\gamma_{\rm max}} \! n_{\rm e}(\gamma) ~ \\
&\times& (1- \rm cos\psi) ~ \mathrm{d} \gamma 
\left( \frac{ \mathrm{d} \sigma }{ \mathrm{d} \epsilon_{\rm{s}} \mathrm{d} \Omega_{\rm{s}} } \right) 
\end{eqnarray}
\citep{dermer09}, where $n_{\rm ph}^{\rm ring}(\epsilon)$ is the external ring photon density per unit volume within the co-moving frame of the blob, $\psi$ is the angle between the scattering electron in the blob and the seed photon from the ring in the co-moving frame of the blob, $  \mathrm{d} \sigma  /  \mathrm{d} \epsilon_{\rm{s}} \mathrm{d} \Omega_{\rm{s}} $ is the differential scattering cross-section, and $\mathrm{d} \Omega$ and $\mathrm{d} \Omega_{\rm{e}}$ are the differential solid angles in the photon and electron directions, respectively.  

In order to evaluate $n_{\rm ph}^{\rm ring}(\epsilon)$ in the co-moving frame of the blob, we follow the method outlined by \citet{dermer93}, and make use of the following Lorentz invariants: $n_{\rm ph}(\epsilon)/\epsilon^2 = n_{\rm ph}(\epsilon^{*})/\epsilon^{*2}$ (where $*$ denotes the host galaxy/stationary ring frame and unprimed notation refers to the co-moving frame of the blob) and $\epsilon^{*}/\epsilon = \Gamma( 1 + \beta \mu )$, where $\Gamma$ is the bulk Lorentz factor of the blob, $\beta \equiv v_{\rm blob}/c ~ = \sqrt{ 1 - \Gamma^{-2} }$ is the blob's velocity in units of $c$, and $\mu$ is the cosine of the angle that the incoming ring photons make with respect to the blob in the co-moving frame of the blob.  The external ring photon density (in $\rm photons ~ cm^{-3} ~ \epsilon^{-1}$) evaluated in the host galaxy frame can be shown to be:
\begin{eqnarray}\nonumber 
n_{\rm ph}(\epsilon^{*}) = \frac{N_{\rm ph}(\epsilon^{*})}{4 \pi d^{2} ~ c} \times ~ \frac{1}{2 \pi} ~ \delta \left( \mu - \frac{ \mu^{*} - \beta }{ 1 - \beta \mu^{*} } \right) 
\end{eqnarray}
\citep{dermer93}, where $d \equiv \sqrt{ r^{2} + z^{2} }$ and $\mu^{*}\equiv \rm cos(\theta) = z / \sqrt{ r^{2} + z^{2} } $, and where $z$ is the distance of the blob from the center of the ring (see Figure \ref{fig2}).  Here, $N_{\rm ph}(\epsilon^{*})$ is the photon production rate (in $\rm photons ~ s^{-1} ~ \epsilon^{-1}$) of the ring due to synchrotron radiation.  If one further assumes that the ring is optically thin; we can write
\begin{eqnarray}\nonumber
N_{\rm ph}(\epsilon^{*}) = \int_{r_{\rm min}}^{r_{\rm max}} \! 4 \pi r ~ h_{\rm o} ~ \dot{n}_{\rm syn}(\epsilon^{*}) ~ \mathrm{d} r ~ ,
\end{eqnarray} 
where $h_{\rm o}$ is the thickness of the ring in the $z$ direction, and where $r_{\rm min}$ and $r_{\rm max}$ denote the inner and outer radii of the ring, respectively (see Figure \ref{fig2}). Here, $\dot{n}_{\rm syn}(\epsilon^{*})$ is the synchrotron photon production rate of the ring (see Equation (\ref{eq1})) evaluated in the host galaxy frame.  It then follows that $n_{\rm ph}^{\rm ring}(\epsilon)$ can be evaluated as follows:
\begin{eqnarray}\nonumber \label{eq5}
n_{\rm ph}^{\rm ring}(\epsilon) &=& \frac{h_{\rm o}}{2 \pi c} ~ \frac{1}{ \Gamma^{2}( 1 + \beta \mu )^{2}} ~ \int_{r_{\rm min}}^{r_{\rm max}} \!  \left( \frac{r}{ r^{2} + z^{2} } \right) \dot{n}_{\rm syn}(\epsilon^{*}) ~ \mathrm{d} r \\
&\times&  \delta \left( \mu - \frac{ \mu^{*} - \beta }{ 1 - \beta \mu^{*} } \right)  ~ .
\end{eqnarray}  

\begin{deluxetable}{lc}
\tablecolumns{2}
\tablewidth{1.0\columnwidth}
\tabletypesize{\normalsize}
\tablecaption{Model Parameters \label{tab1}}
\startdata
\tableline
\noalign{\smallskip}
\tableline
\noalign{\smallskip}
Global Parameter & Value \\
\tableline
\noalign{\smallskip}
$\gamma_{\rm{min}}$ & $2.0 \times 10^{3}$ \\
\noalign{\smallskip}
$\gamma_{\rm{max}}$ & $1.0 \times 10^{4}$ \\
\noalign{\smallskip}
$Z$ & 0.361 \\
\noalign{\smallskip}
$\theta_{\rm{obs}}$ & $1.4^{\circ}$ \\
\noalign{\smallskip}
$\rm{Baseline} ~ \rm{Flux}_{~ \rm{optical}}$ & 1.5 (mJy) \\
\noalign{\smallskip}
$\rm{Baseline} ~ \rm{ Flux}_{~ \gamma-\rm{ray}}$ & 6.5 ($10^{-7} \rm{phot} ~ \rm{cm}^{-2} ~ \rm{s}^{-1}$) \\
\noalign{\smallskip}
\tableline
\noalign{\smallskip}
Blob Parameter & Value \\
\tableline
\noalign{\smallskip}
$r_{\rm{blob}}$ & 0.01 (pc)  \\
\noalign{\smallskip}
$\Gamma_{\rm{initial}}$ & 4 \\
\noalign{\smallskip}
$\Gamma_{\rm{final}}$ & 19 \\
\noalign{\smallskip}
$z_{\rm{initial}}$ & -0.1 (pc) \\
\noalign{\smallskip}
$z_{\rm{final}}$ & 0.1 (pc) \\
\noalign{\smallskip}
$s_{\rm{blob}}$ & 4.0 \\
\noalign{\smallskip}
$B_{\rm{blob}}$ & 0.03 (G) \\
\noalign{\smallskip}
$P_{\rm{inj}}$ & $1.65 \times 10^{43}$ ($\rm{ergs} ~ s^{-1}$) \\
\noalign{\smallskip}
\tableline
\noalign{\smallskip}
Ring Parameter & Value \\
\tableline
\noalign{\smallskip}
$z_{\rm{ring}}$ & 0.0 (pc) \\
\noalign{\smallskip}
$r_{\rm{min}}$ & 0.09 (pc) \\
\noalign{\smallskip}
$r_{\rm{max}}$ & 0.18 (pc) \\
\noalign{\smallskip}
$h_{\rm{o}}$ & 0.018 (pc) \\
\noalign{\smallskip}
$s_{\rm{ring}}$ & 4.0 \\
\noalign{\smallskip}
$B_{\rm{ring}}$ & 0.12 (G) \\
\enddata
\end{deluxetable}

A head-on approximation is adopted, in which the scattered photon direction is set equal 
to the electron direction in the blob frame, which then results in the scattering angle: $\rm cos\psi = \mu \mu_{\rm s} + \sqrt{1-\mu^{2}} \sqrt{1-\mu_{\rm s}^{2}} ~ \rm cos\phi$ \citep{dermer09}, where $\mu_{\rm s} = ( \mu_{\rm obs} - \beta)/(1 - \beta \mu_{\rm obs})$,  and $\mu_{\rm obs} = \rm cos(\theta_{\rm obs})$, where $\theta_{\rm obs}$ is the angle between the jet axis and our line of sight (see Figure \ref{fig2}), and $\phi$ is the azimuthal angle about the jet axis.  In order to keep the numerical integration to a minimum, a delta-function approximation of the differential scattering cross-section in the Thomson regime is adopted:
\begin{equation} \label{eq6}
\frac{ \mathrm{d} \sigma }{ \mathrm{d} \epsilon_{\rm{s}} \mathrm{d} \Omega_{\rm{s}} } = \sigma_{\rm T} \delta( \Omega_{\rm s} - \Omega_{\rm e} ) \delta( \epsilon_{\rm s} - \gamma \bar{\epsilon} ) H( 1 - \bar{\epsilon} )  
\end{equation}
\citep{dermer09}, where $\Omega_{\rm s}$ refers to the solid angle subtended by the scattered photons, $\bar{\epsilon} \equiv \gamma \epsilon ( 1 - \rm cos\psi)$, and $H( 1 - \bar{\epsilon} )$ is a Heaviside function given by:
\begin{equation} \label{eq7}
H( 1 - \bar{\epsilon} ) = \left\{ \begin{array}{ll}
   0 & \textrm{for 
$ ( 1 - \bar{\epsilon} ) < 1$}\\
\\
\frac{1}{2} & \textrm{for 
$ ( 1 - \bar{\epsilon} ) = 1$}\\
\\
   1 & \textrm{for 
$( 1 - \bar{\epsilon} ) > 1$}~.
\end{array} \right.
\end{equation}
As discussed in \citet{dermer93}, Equation (\ref{eq6}) is a valid approximation of the scattering cross-section provided $\bar{\epsilon} \leq 1$.  The delta-function $\delta( \epsilon_{\rm s} - \gamma \bar{\epsilon} )$ in Equation (\ref{eq6}) implies $\bar{\epsilon} = \epsilon_{\rm s}/\gamma$.  The dimensionless scattered photon energy ($\epsilon_{\rm s}$) is defined as $\epsilon_{\rm s} = h \nu_{\rm s} / m_{\rm e} c^{2}$.  The orphan flare (presented in Figure \ref{fig1}) we model with our code occurs at an observational frequency of $\nu_{\rm obs} \sim 10^{23} ~ \rm Hz$.  Due to the relativistic motion of the blob, the scattered photon frequency ($\nu_{\rm s}$) in the co-moving frame of the blob is related to the observed frequency by $ \nu_{\rm s} = \nu_{\rm obs}/D$, where $D$ is a relativistic Doppler boosting factor given by $\rm{D} \equiv 1 / \Gamma ( 1 - \beta ~ \rm cos \theta_{\rm obs} )$.  With these quantities defined, we compute the limiting values of $\bar{\epsilon}$ for the range of scattered photon and electron energies modeled in our code (see Table \ref{tab1}), and we find: $0.003 < \bar{\epsilon} < 0.05$.  This range of $\bar{\epsilon}$ falls well within the Thomson regime.      
In the future, the code will incorporate the effects of the reduced Klein--Nishina scattering cross-section at higher energies.  Since $\oint \! \mathrm{d} \Omega \equiv \int \mathrm{d} \mu \int \mathrm{d} \phi$, Equations (\ref{eq4})--(\ref{eq6}) can be combined to yield a simplified version of the external Compton photon production rate:
\begin{eqnarray}\nonumber \label{eq8}
\dot{n}_{\rm{ec}}(\epsilon_{\rm{s}}) &=&
\left( \frac{\sigma_{\rm{T}} h_{\rm{o}}}{2\pi} \right)  
\int_{0}^{2 \pi} \! \mathrm{d} \phi 
\int_{r_{\rm{min}}}^{r_{\rm{max}}} \! \mathrm{d} r 
\int_{\gamma_{\rm{min}}}^{\gamma_{\rm{max}}} \! \mathrm{d} \gamma \\
&\times&\nonumber  
\frac{1}{ \Gamma^{2}( 1 + \beta \mu )^{2}}  ~ \left( \frac{r}{r^2 + z^2} \right) ~ n_{\rm{e}}(\gamma)  \\
&\times&
 \dot{n}_{\rm{syn}}(\epsilon^{*}) ~ ( 1- \rm{cos} \psi  ) ~ H( 1 - \bar{\epsilon} ) ~ ,
\end{eqnarray} 
where now $\epsilon =  \epsilon_{\rm{s}} / \gamma^{2}( 1 - \rm{cos}\psi )  $ and $\mu = (\mu^{*}-\beta)/(1-\beta \mu^{*})$.  

\begin{figure}
\epsscale{1.15}
\plotone{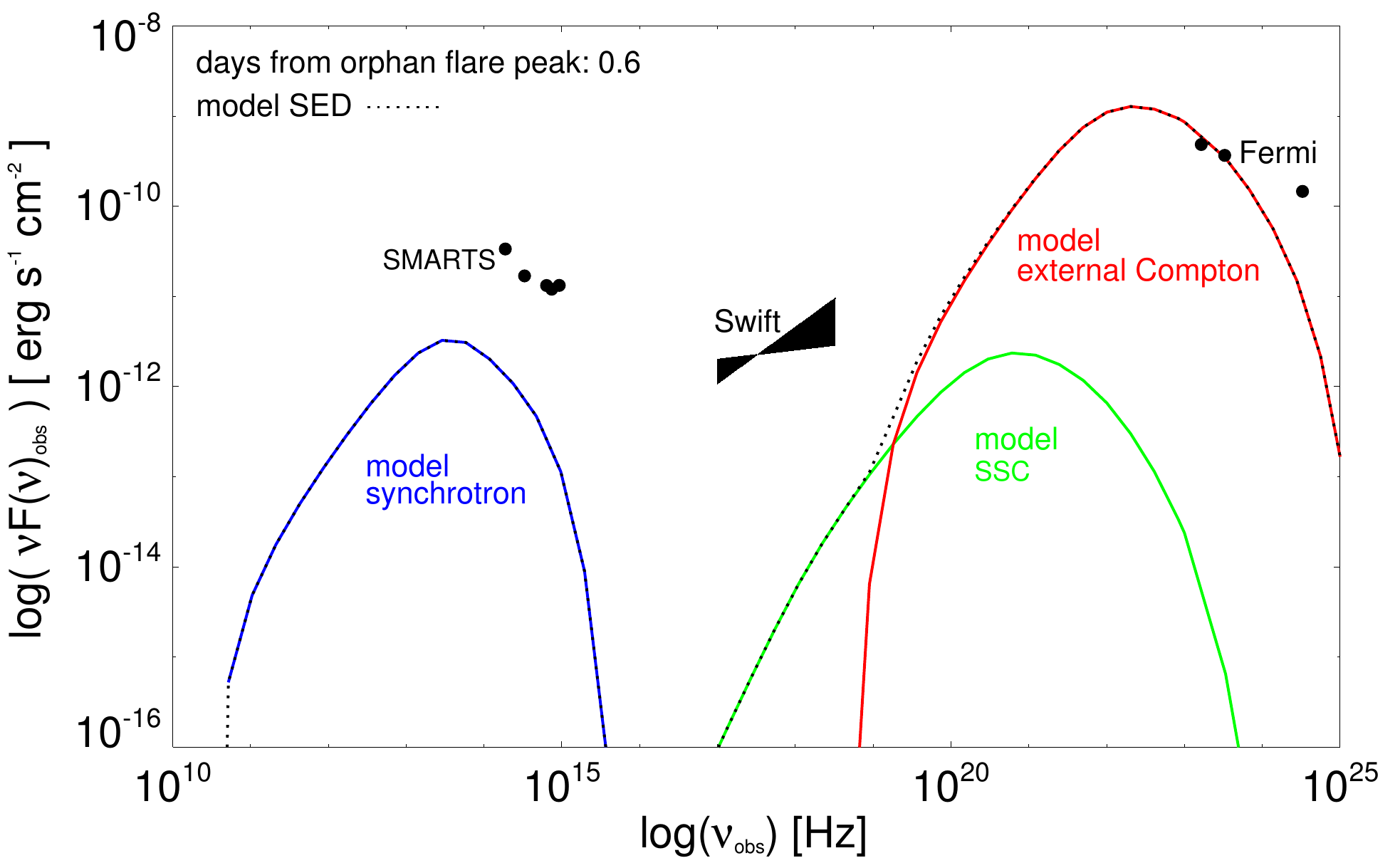}
\caption{\label{fig3}SED produced by the model during the orphan $\gamma$-ray flare.  The SED is composed of synchrotron (blue), synchrotron self-Compton (green), and external-Compton (red) emission components.  Multi-wavelength data obtained by the SMARTS, \textit{Swift}, and \textit{Fermi} telescopes during the epoch of the orphan $\gamma$-ray flare presented in Figure \ref{fig1} are plotted in black for comparison. }
\end{figure}

The distribution of electrons in the blob is evolved each time step by solving the Fokker--Planck equation:
\begin{equation} \label{eq9}
\frac{ \partial n_{\rm{e}}(\gamma) }{ \partial t } = -b ~ m_{\rm{e}}c^{2} ~ \frac{ \partial }{ \partial \gamma }( ~ \gamma^{2} n_{\rm{e}}(\gamma) ~ ) + q\gamma^{2} 
\end{equation}
\citep{kardashev62}, where $b$ is the electron radiative cooling rate in $\rm ergs^{-1} ~ s^{-1}$, and $q$ is the electron injection rate per unit dimensionless energy in $\rm cm^{-3} ~ s^{-1}$.  In general, $b = b_{\rm syn} + b_{\rm ssc} + b_{\rm ec}$;  however, here we assume that $b \approx b_{\rm syn} +  b_{\rm ec}$ because ssc is only a minor component in the SEDs produced by the runs of our code presented here (see Figure \ref{fig3}).  In the future we intend to carry out a more sophisticated calculation that includes a proper treatment of the ssc cooling, which, as pointed out by \citet{zacharias12}, can strongly influence the evolution of $n_{\rm{e}}(\gamma)$ in time-dependent calculations of the non-thermal emission from some blazars.  For simplicity, we neglect aging in the ring electrons.  The synchrotron cooling rate for the blob electrons is given by:
\begin{equation} \label{eq10}
b_{\rm syn} = \frac{4}{3}  c \sigma_{\rm{T}} \left( \frac{B^{2}}{8\pi m_{\rm{e}}^{2} c^{4}} \right) 
\end{equation}
\citep{dermer09}.  The external Compton cooling rate is given by:
\begin{equation} \label{eq11}
b_{\rm ec} = \frac{4}{3} c \sigma_{\rm T} \left( \frac{1}{m_{\rm e}c^2} \right) \int_{0}^{\gamma_{\rm min}} \! n_{\rm ph}^{\rm ring}(\epsilon) ~ \mathrm{d} \epsilon 
\end{equation} 
\citep{dermer09}, where the upper limit of the integration ensures that the cooling rate is evaluated within the Thomson regime.

The injection rate $q$ is computed by defining an injection power:
\begin{eqnarray} \label{eq12}
P_{\rm inj} &=& \int_{\gamma_{\rm min}}^{\gamma_{\rm max}} \! ( \gamma m_{\rm e} c^{2} )( q \gamma^{-s} )\left( \frac{4}{3} \pi r_{\rm blob}^{3} \right) \mathrm{d} \gamma \\
\rightarrow q &=& \frac{ (2-s) P_{\rm inj} }{ ( m_{\rm e} c^{2} )\left( \frac{4}{3} \pi r_{\rm blob}^{3} \right)( \gamma_{\rm max}^{ 2-s } - \gamma_{\rm min}^{ 2-s }  ) } ~ (\rm{for} ~ s > 2) \nonumber.
\end{eqnarray}
The injection power parameterizes a physical mechanism at work within the blob, for example, turbulence or diffusive shock acceleration, that continuously rejuvenates the aging blob electron distribution.  This injection power is a free parameter within our model and has been set at $P_{\rm inj} \sim 10^{43} ~ \rm ergs ~ s^{-1}$ in the calculations presented here.

Equation (\ref{eq9}) has the following analytical solution:
\begin{equation} \label{eq13}
n_{\rm{e}}(\gamma,t) =  \left\{ \begin{array}{ll}
    n_{ \rm o } \gamma^{-s}  & \textrm{for 
$ \xi = 0$}\\
\\ 
n_{ \rm o } \gamma^{-s}( 1 - \xi  )^{ s-2 } \nonumber \\ + \frac{ q \gamma^{-(s+1)} }{ m_{\rm e} c^{2} b ( s-1 )  } \left[ 1 - ( 1 - \xi  )^{ s-1 } \right]   & \textrm{for 
$ \xi < 1$}\\
\\
   \frac{ q \gamma^{-(s+1)} }{ m_{\rm e} c^{2} b ( s-1 )  }  & \textrm{for 
$ \xi \ge 1$}~,
\vspace{5mm}
\end{array} \right. 
\vspace{5mm}
\end{equation} 
where $\xi \equiv b t \gamma m_{\rm e} c^{2}$ \citep{kardashev62}.  The initial relativistic electron number density, $n_{\rm o} $, of the blob is determined by assuming equipartition between the magnetic and electron energy densities within the blob:  
\begin{eqnarray}\nonumber \label{eq14}
\left( \frac{B_{\rm blob}^{2}}{8 \pi} \right) &=& \int_{\gamma_{\rm min}}^{\gamma_{\rm max}} \! ( \gamma m_{\rm e} c^{2} )(  n_{\rm o} \gamma^{-s} ) \mathrm{d} \gamma \\
\rightarrow n_{\rm o} &=& \frac{ \left( \frac{ B_{\rm blob}^{2} }{ 8 \pi } \right) \frac{ 2-s }{ m_{e} c^2 } }{ \gamma_{\rm max}^{ 2-s } - \gamma_{\rm min}^{ 2-s } } ~ (\rm{for} ~ s > 2).
\end{eqnarray} 

After the photon production rates have been computed for a range of photon energies at each time step, the 
flux is computed in the observer's frame by multiplying the sum of the production rates ($ \dot{n}_{\rm tot} =  \dot{n}_{\rm syn} +  \dot{n}_{\rm ssc} +  \dot{n}_{\rm ec}$) by the volume
of the blob, the energy of the photons, and a $( \mathrm{d} \epsilon /  \mathrm{d} \nu )$ conversion factor:
\begin{eqnarray}\nonumber \label{eq15}
\nu F(\nu)_{\rm obs} &=& \frac{ 1 }{ 4 \pi d_{\rm L}^{2} } \times \frac{ \rm{D}^{4} }{ 1 + Z } \times \dot{n}_{\rm tot}(\epsilon) \times \frac{4}{3} \pi r_{\rm blob}^{3} \\
&\times&  ~  \frac{ ( h \nu )^{2} }{ m_{\rm e} c^{2} },
\end{eqnarray}
\noindent
where $Z$ is the redshift, $d_{\rm L} $ is the luminosity distance, and $\rm{D}$ is a relativistic Doppler boosting factor, given by $ \rm{D} \equiv 1 / \Gamma ( 1 - \beta ~ \rm cos \theta_{\rm obs} )$.

\bigskip

\section{Numerical Simulation}

The numerical calculations were carried out using the BLAZE code (people.bu.edu/nmacdona/Research.html).  A spherical blob of radius $r_{\rm{blob}}= 0.01 ~ \rm{pc}$ accelerates from a bulk Lorentz factor of $\Gamma_{\rm initial} = 4$ to $\Gamma_{\rm final} = 19$, after which it continues to propagate at constant speed along the spine of the jet.  The parameters governing the blob acceleration have been chosen to reflect the acceleration derived for the emission knot associated with the orphan $\gamma$-ray flare in PKS 1510$-$089 (see Figure \ref{fig1}) based on the changing rate of rotation of the observed optical polarization vector (\citealt{marscher08}, \citeyear{marscher10}).  The blob begins its acceleration toward the ring (located at $z_{\rm ring} = 0 ~ \rm pc$ in the simulation) at $z_{\rm intial} \sim -0.1 ~ \rm pc$ and reaches $\Gamma_{\rm final}$ at $z_{\rm final} \sim 0.1 ~ \rm pc$ after passing through the ring.  The blob is tracked out to a distance of $z_{\rm end} \sim 0.4 ~ \rm pc$, after which the simulation is terminated.  Although in actuality the sheath (and by extension the ring) will have some relatively low velocity along the jet, for the purposes of the numerical simulation, we treat the ring as stationary.

The dimensions of the ring used in the simulation, namely, its minimum ($r_{\rm min}$) and maximum ($r_{\rm max}$) radius and its thickness ($h_{o}$) (see Figure \ref{fig2}), are listed in Table \ref{tab1}.  The size of the sheath (and by extension the ring) is thought to be a substantial fraction of the cross-sectional width of the jet (\citealt{zavala04}; \citealt{walg13}) and the values chosen for $r_{\rm min}$ and $r_{\rm max}$ reflect the width of the jet in PKS 1510$-$089.  Since we are positing that the ring represents a shocked segment of a continuous jet sheath (see Figure \ref{fig9} below), the thickness of the ring ($h_{o}$) in our model reflects the length scale over which particle acceleration acts on a small section of the sheath, perhaps as it propagated through a standing shock upstream of its current location within the jet.  While this parameter is the least constrained in our model, $h_{o}$ simply scales the external ring photon density up or down, depending on our choice of its value (see Equation (\ref{eq5})).    

\citet{nalewajko14} make a case against the ability of a sheath of plasma surrounding the jet to provide sufficient numbers of seed photons to produce the orphan flares seen in PKS 1510$-$089 (i.e., Figure \ref{fig1}) without the sheath itself outshining the observed flux.  We find no conflict, however, between our model and observations of this blazar.  As discussed in their paper, the observed luminosity of the sheath can be shown to be (their Equation (C3)):
\begin{eqnarray}\nonumber \label{eq16}
L_{\rm sh,obs} \simeq 2.7 &\times& 10^{47} \rm erg ~ s^{-1} \left( \frac{ r }{ 10 \rm pc } \right)^{2} \left( \frac{ \Gamma }{ 20 } \right)^{-2} \\
                                               &\times& \Gamma_{\rm sh}^{4} ( \Gamma_{\rm sh} \theta_{\rm sh} )^{2} ,
\end{eqnarray} 

\noindent where $r$ is the distance along the jet, $\Gamma$ is the blob's bulk Lorentz factor, and $\Gamma_{\rm sh}$ and $\theta_{\rm sh}$ are the bulk Lorentz factor and opening angle of the sheath, respectively.  As \citeauthor{nalewajko14} point out, the strong dependence of $L_{\rm sh,obs}$ on $\Gamma_{\rm sh}$ means that for even mildly relativistic sheaths the observed luminosity can become quite large.  The ring of sheath plasma in our model, however, is located upstream of the radio core in PKS 1510$-$089.  In particular, the radio knot correlated with the orphan flare was traveling down the jet at $\sim 0.3 ~ \rm{pc} ~\rm{day}^{-1}$ in our frame \citep{marscher10}.  From Figure \ref{fig1}, it is apparent that the onset of the orphan flare preceded the large optical flare associated with the knot passing the radio core by about 20 days.  Assuming that the knot moved at this observed rate along the jet for those 20 days, that would place our ring $6 ~ \rm{pc}$ upstream of the radio core.  If we further assume that the radio core is at roughly a scale of $\sim 10 ~ \rm{pc}$ from the central engine, that would place the ring at roughly a distance of $r \sim 4 ~ \rm pc$ along the jet.  This locates our emission region between the two commonly considered regions of $\gamma$-ray production in blazar jets, namely, the near-dissipation region ($r \sim 1 ~ \rm pc$), where the external photon fields are dominated by emission from a dusty torus and the BLR, and the far-dissipation region ($r \gtrsim 10 ~ \rm pc$) near the core seen at 43 GHz.  At this distance, utilizing the dimensions of our ring as corresponding to the jet sheath, we compute a sheath opening angle of $\theta_{\rm sh} \sim 1.29^{\circ}$ (where $\theta_{\rm sh} \equiv \theta_{\rm max} - \theta_{\rm min}$).  Here, $\theta_{\rm min}$ and $\theta_{\rm max}$ refer to the inner and outer angles subtended by the ring relative to the jet apex, which we take to be $4 ~ \rm pc$ upstream of the ring--see Figure \ref{fig9}--and can be shown to be: $\theta_{\rm min} = \rm tan^{-1}( r_{\rm min}/4 ~ \rm pc ) \sim 1.29^{\circ}$ and $\theta_{\rm max} = \rm tan^{-1}( r_{\rm max}/4 ~ \rm pc ) \sim 2.58^{\circ}$, respectively.  We point out that this value for the sheath opening angle is in rough agreement with the sheath opening angle of $\theta_{\rm sh} \sim 4.6^{\circ}$ used by \citet{darcangelo09} to explain variability in the polarization of the blazar OJ 287.  If we further assume that the sheath at this scale is only mildly relativistic ($\Gamma_{\rm sh} \sim 2$) and take our final value for the bulk Lorentz factor of the blob ($\Gamma \sim 19$), Equation (\ref{eq16}) yields $L_{\rm sh,obs} \sim 1.5 \times 10^{45} \rm ~ erg ~ s^{-1}$, which is two orders of magnitude smaller than the observed $\gamma$-ray luminosity for PKS 1510$-$089 during this flaring epoch ($L_{\gamma} \sim 5.4 \times 10^{47} \rm erg ~ s^{-1}$) and is roughly a factor of 3.6 smaller than the observed synchrotron luminosity ($L_{\gamma}/L_{\rm syn} \sim 100 \rightarrow L_{\rm syn} \sim 5.4 \times 10^{45} \rm erg ~ s^{-1}$). 

The blob and ring electron populations are both modeled as identical power-law distributions, $n_{e}(\gamma)$, with identical power-law indices $s$ (listed in Table \ref{tab1}), in rough agreement with the observed $\gamma$-ray photon index of $\Gamma = 2.48$ ($\alpha = \Gamma - 1 ; s = 2 \alpha + 1$) reported in \citet{marscher10}.  By setting the power-law index to this value, we are making the assumption of an unbroken power-law of electron energies in both regions of our model.  The range of electron energies over which we assume the power-law applies ($\gamma_{\rm min}$ through $\gamma_{\rm max}$) is listed in Table \ref{tab1}.   

Constant magnetic fields are taken to permeate both the blob ($B_{\rm{blob}}$) and the ring ($B_{\rm{ring}}$) with the values used in the simulation listed in Table \ref{tab1}.  As discussed in \citet{nishikawa13}, the jet sheath is thought to have a higher electron density than the spine.  Since we assume equipartition, this implies that $B_{\rm{ring}} > B_{\rm{blob}}$.  In addition, a lower magnetic field within the blob is needed to keep the radiative losses experienced by the blob electrons out of the fast cooling regime (see Equation (\ref{eq13})); however, the value used in our simulation ($B_{\rm{blob}} = 0.03 ~ G$) is at odds with the observed value of $B_{\rm{jet}} \sim 0.1-1 ~ G$ for PKS 1510$-$089 reported in \citet{marscher10}.  This difference in field strength might be reconciled by the fact that, as discussed above, we treat the ring as being located upstream of the radio core in PKS 1510$-$089.  It is consistent for the magnetic fields in the jet upstream of the core to be smaller if indeed the core at 43 GHz is associated with a recollimation shock that amplifies the field strength through shock compression (see Figure \ref{fig9} below).   We also approximate that the magnetic fields within the blob and the ring do not vary appreciably on the timescale required for the blob to pass through the ring ($\sim 100 ~ \rm days$ in the AGN frame).  This requires that the opening angle of the jet $< 10^{\circ}$, which agrees with $\theta \sim 1^{\circ}$ found by \citet{jorstad05}.   

The numerical integrations were performed using linear grids of 15 radial and azimuthal zones and logarithmic grids of 60 intervals in $\epsilon$ and in $\gamma$.  In order to convert the resultant flux into the observer's frame, the jet is taken to lie at an inclination angle of $\theta_{\rm obs} = 1.4^{\circ}$ \citep{jorstad05}, with a redshift of $Z = 0.361$ \citep{thompson90}, and a luminosity distance of $d_{\rm L} = 1913 \rm ~ Mpc$.  These values are based on the observed properties of PKS 1510$-$089 and yield a projected scale of $5.0 ~ \rm pc ~mas^{-1}$.  To better fit the observed light curves presented in Figure \ref{fig1}, we include a baseline flux in each band, which is added to the variable emission produced by the model.  These baseline fluxes, listed in Table \ref{tab1}, reflect the more slowly varying emission from the jet plasma and are approximated to be constant.  Finally, the time step is chosen so that 24 samplings of the SED are made by the end of the simulation of the blob passing through the ring.

\section{Results}

Figure \ref{fig3} shows an SED produced by the model during the peak of the simulated orphan $\gamma$-ray flare.  The three non-thermal emission components included in the calculation are differentiated by color.  The model broadly reproduces the observed optical and $\gamma$-ray peaks of PKS 1510$-$089's SED (shown in black), but does not reproduce the observed optical-ultraviolet excess (known as the big blue bump), nor the observed X-ray flux levels.  The former spectral feature is believed to be associated with emission from the accretion disk surrounding the central engine, while the latter may be ssc emission from regions farther down the jet (as indicated by the weak correlation between the X-ray and $\gamma$-ray light curves; see \citealt{marscher10}).  Neither of these components are included in the calculation, hence these spectral features are absent from the synthetic SEDs produced by the model.

\begin{figure}
\epsscale{1.17}
\plotone{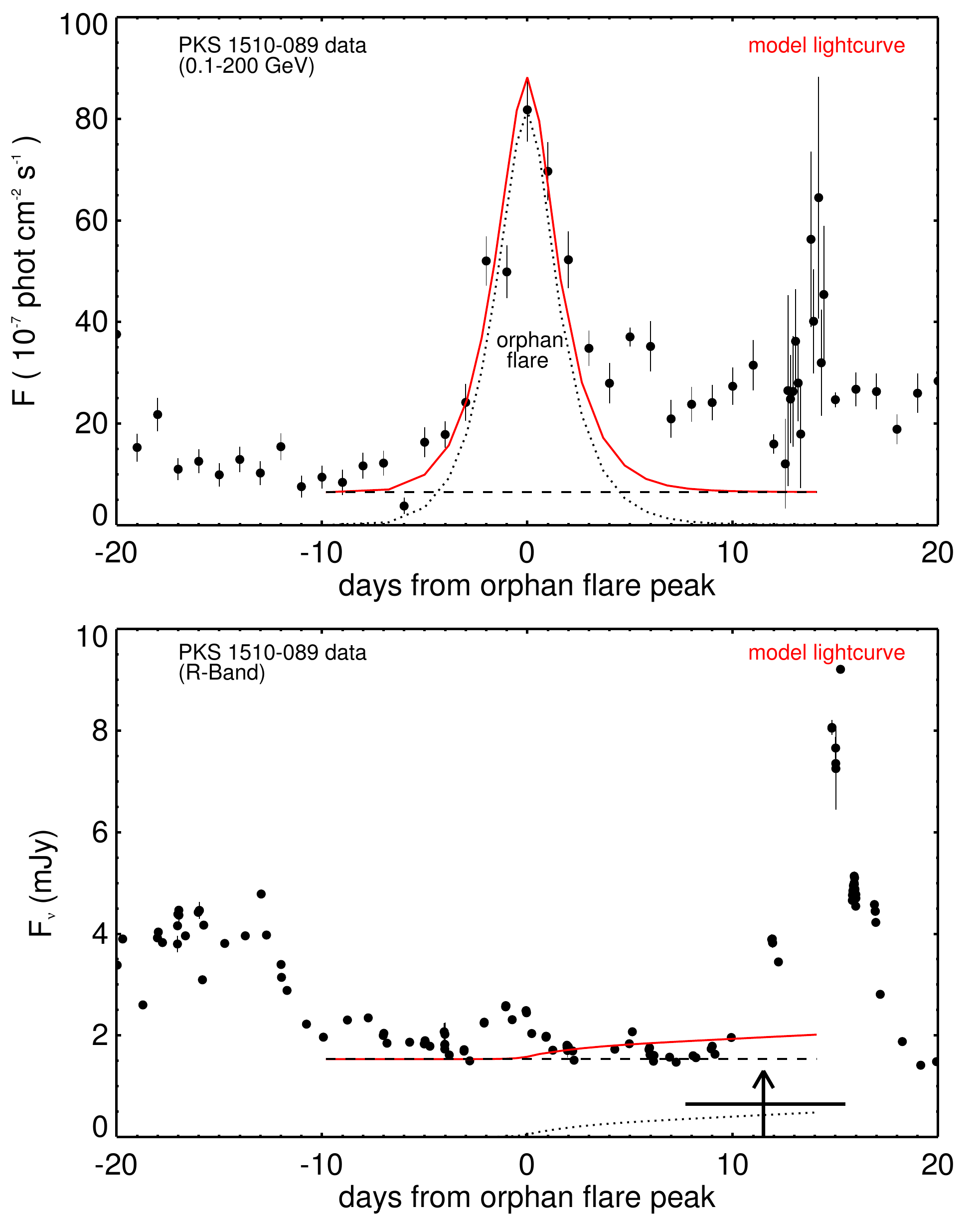}
\caption{\label{fig4}PKS 1510$-$089 $\gamma$-ray (upper panel) and optical (lower panel) light curves from 2009 (black circles)--a smaller segment of Figure \ref{fig1}.  The model light curves are overlaid (in red) and consist of the superposition of the model baseline level of flux in each band (the dashed black line) and the variable emission produced by the model (the dotted black line).  As with Figure \ref{fig1}, the vertical arrow in the lower panel marks the time when a superluminal knot passed through the 43 GHz core of PKS 1510$-$089, with the horizontal bar representing the uncertainty in this time.}
\end{figure}

The resultant $\gamma$-ray and optical light curves produced by the blob's passage through the ring are shown in Figure \ref{fig4} (highlighted in red) overlaid upon the observed data from \citet{marscher10}.  In particular, the baseline flux level (as discussed in \S3) is demarcated by the black dashed line in each plot, whereas the variable emission produced by the model is indicated by the black dotted line.  The superposition of these two lines in each plot creates the synthetic light-curves shown in red.  The synthetic light curves reproduce, in general, the observed orphan $\gamma$-ray flare without exceeding the observed optical flux during the flare.  The lack of a strong optical flare demonstrates the model's key ability to produce orphan $\gamma$-ray flares.  The peak flux in the simulated $\gamma$-ray flare occurs when the blob is at a distance of $z = 0.08 ~ \rm{pc}$ downstream of the ring.  This is a result of the interplay between the Doppler boost gained by the relativistic motion of the blob, the external ring photon density within the co-moving frame of the blob, and the effects of radiative cooling.  It should be pointed out that there is a small optical ``bump'' in our synthetic light curve (lower panel of Figure \ref{fig4}).  This gradual increase in emission is a result of the Doppler boost gained by the blob's acceleration along the spine of the jet.  

\begin{figure}
\epsscale{1.17}
\plotone{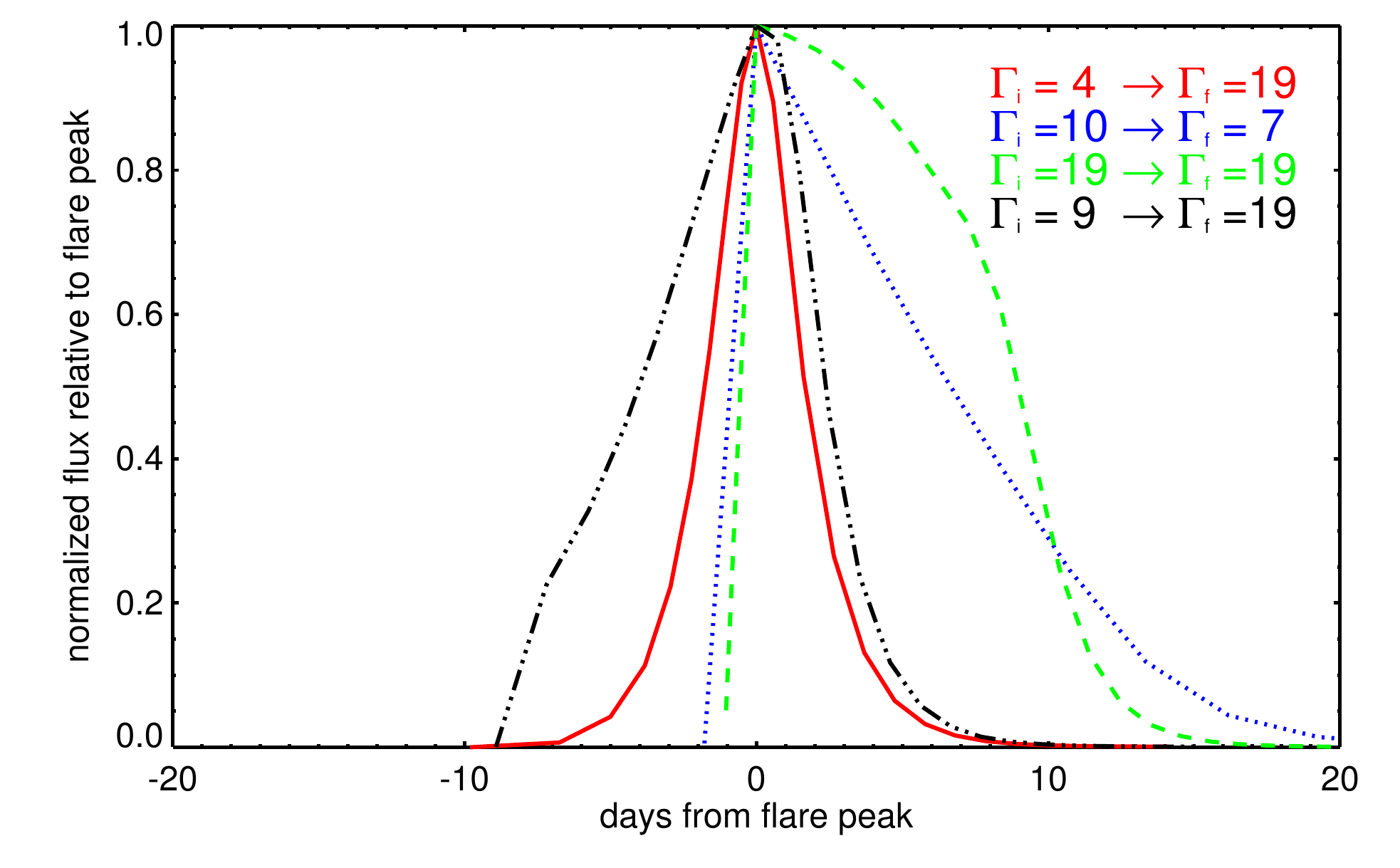}
\caption{\label{fig5}Comparative light curves highlighting the effect different motions of the blob along the spine have on the orphan flares produced by this model.  In all cases, the flux for each flare has been normalized to the flux of that flare's peak.  For comparison, the fit to the orphan flare shown in Figure \ref{fig4} is reproduced in red.  The blob's initial ($\Gamma_{\rm{initial}}$) and final ($\Gamma_{\rm{final}}$) bulk Lorentz factor for each run are listed in the upper right portion of the plot (color-coded to the corresponding flare).  It is apparent from the blue curve that if the blob decelerates as it approaches the ring, the resultant flare has a steeper onset time in comparison with its decay time.  A similar effect (albeit with a distinct profile) is produced if the blob moves through the ring at constant speed (green curve).  In contrast to the relatively symmetric flare profile (shown in red), if the blob approaches the ring with a greater velocity (black curve), the resultant flare has a more gradual onset.}
\end{figure}

The model's ability to produce $\gamma$-ray flares with steep rise and decay times (as illustrated in the upper panel of Figure \ref{fig4}) reflects the anisotropic source of seed photons used in the simulation (i.e., the ring).  This is in contrast to other structured jet models that invoke isotropic seed photon fields from a continuous jet sheath.  The shape of the orphan $\gamma$-ray flare shown in Figure \ref{fig4} is very sensitive to the nature of the blob's acceleration toward and through the ring.  To further explore the effect the blob's motion along the jet has on the orphan $\gamma$-ray flares produced by this model, the parameters governing the blob's acceleration along the spine of the jet were varied, namely, the blob's initial ($\Gamma_{\rm{initial}}$) and final ($\Gamma_{\rm{final}}$) bulk Lorentz factors.  In Figure \ref{fig5}, a series of comparative light curves is shown illustrating the effect that different blob accelerations (or lack of acceleration) have on the orphan flares produced by this model.  It is apparent from Figure \ref{fig5} that by varying the blob's acceleration through the ring, this model is able to produce orphan flares both with steeper and more gradual onsets and decays.  The differences in these light curves highlight the potential to use the observed shape of orphan $\gamma$-ray flares to infer the nature of the acceleration of the emission region through blazar jets upstream of the 43 GHz core.

\section{Polarimetric Signature of a Jet Sheath}

\begin{figure*}
  \setlength{\abovecaptionskip}{-6pt}
  \begin{center}
    \scalebox{0.85}{\includegraphics[width=2.0\columnwidth,clip]{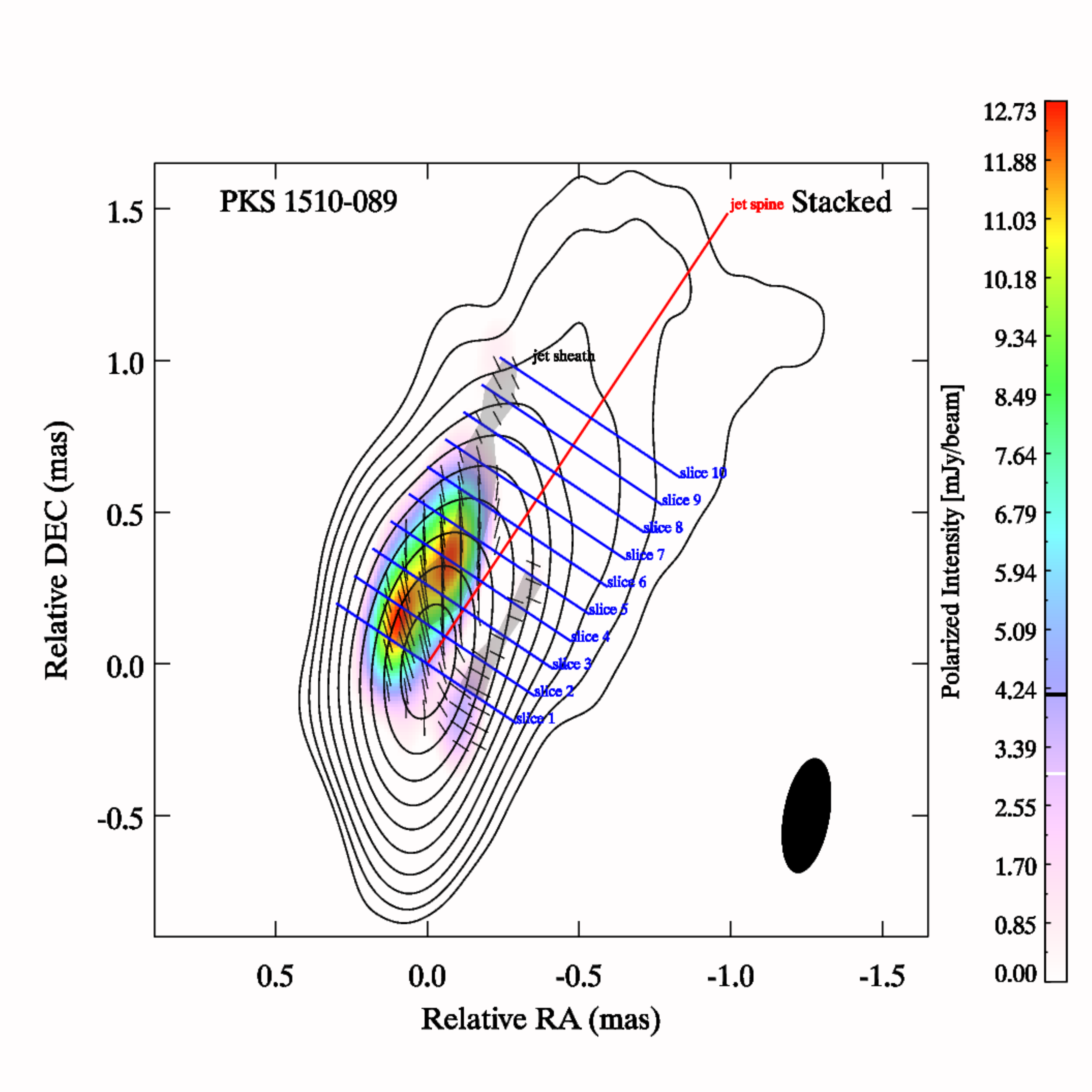}}
  \end{center}
  \caption{\label{fig6}A stacked map of 43 GHz images of PKS 1510$-$089 spanning twenty epochs of observation from 2008 to 2013.  The black contours correspond to total intensity (contour levels are (mJy beam$^{-1}$): 2.3 4.6 9.2 18.4 36.8 73.6 147.2 294.4 588.8 1177.6), whereas the underlying color scheme corresponds to polarized intensity (see color bar to the right for the flux levels) with the EVPAs denoted by black line segments.  All images have been convolved with a Gaussian beam, shown on the lower right of the stacked map.  An arbitrary jet spine has been plotted in red, across which ten transverse slices through the data are taken (shown in blue).  The profiles of the emission parameters along the first slice are shown in Figure \ref{fig7}.  The shaded gray regions in the above map highlight the nominal location of the jet sheath of PKS 1510$-$089 as determined by the procedure discussed in \S5 and highlighted in Figure \ref{fig8}.}
\end{figure*}

The presence of a sheath of plasma surrounding the spine of a blazar jet is a central component to our model of orphan $\gamma$-ray flares.  Several authors have discussed what the observational signature of a jet sheath within a blazar would be (see \citealt{pushkarev05} and references therein).  They posit that a jet sheath represents a shear layer between the plasma of the jet and the ambient medium into which the jet propagates.  In this scenario, the initially helical or chaotic magnetic field lines of the outer layers of the jet are ``stretched'' out due to the velocity gradient between the jet and the ambient medium.  This shear, if it exists, would result in a buildup of magnetic field on the outer edges of the jet, with the orientation of that field aligned nearly parallel to the jet boundary \citep{wardle94}.  The direction of the magnetic field within a blazar is inferred from the orientation of the EVPAs measured by the VLBA, which, barring the effects of opacity and relativistic aberration \citep[see][]{lyutikov05}, are assumed to be perpendicular to the projection of the magnetic field onto the plane of the sky.  Therefore, the observational signature of a jet sheath would be an increase in the fractional polarization toward the edges of the jet, with the EVPA perpendicular to the jet boundary, and therefore roughly perpendicular to the jet axis.  We point out that an alternative interpretation (as discussed in \citealt{lyutikov05}) is that the EVPA being perpendicular to the jet boundary is the result of the jet flow carrying a large scale helical magnetic field.  A measure of the Faraday rotation gradient, if it exists, \citep[see][]{gabuzda13} across the width of the jet of PKS 1510$-$089, would potentially be able to distinguish between these two scenarios.  We adopt the prior interpretation for this paper.    

\begin{figure}
  \setlength{\abovecaptionskip}{-6pt}
  \begin{center}
    \scalebox{1.0}{\includegraphics[width=1.4\columnwidth,clip]{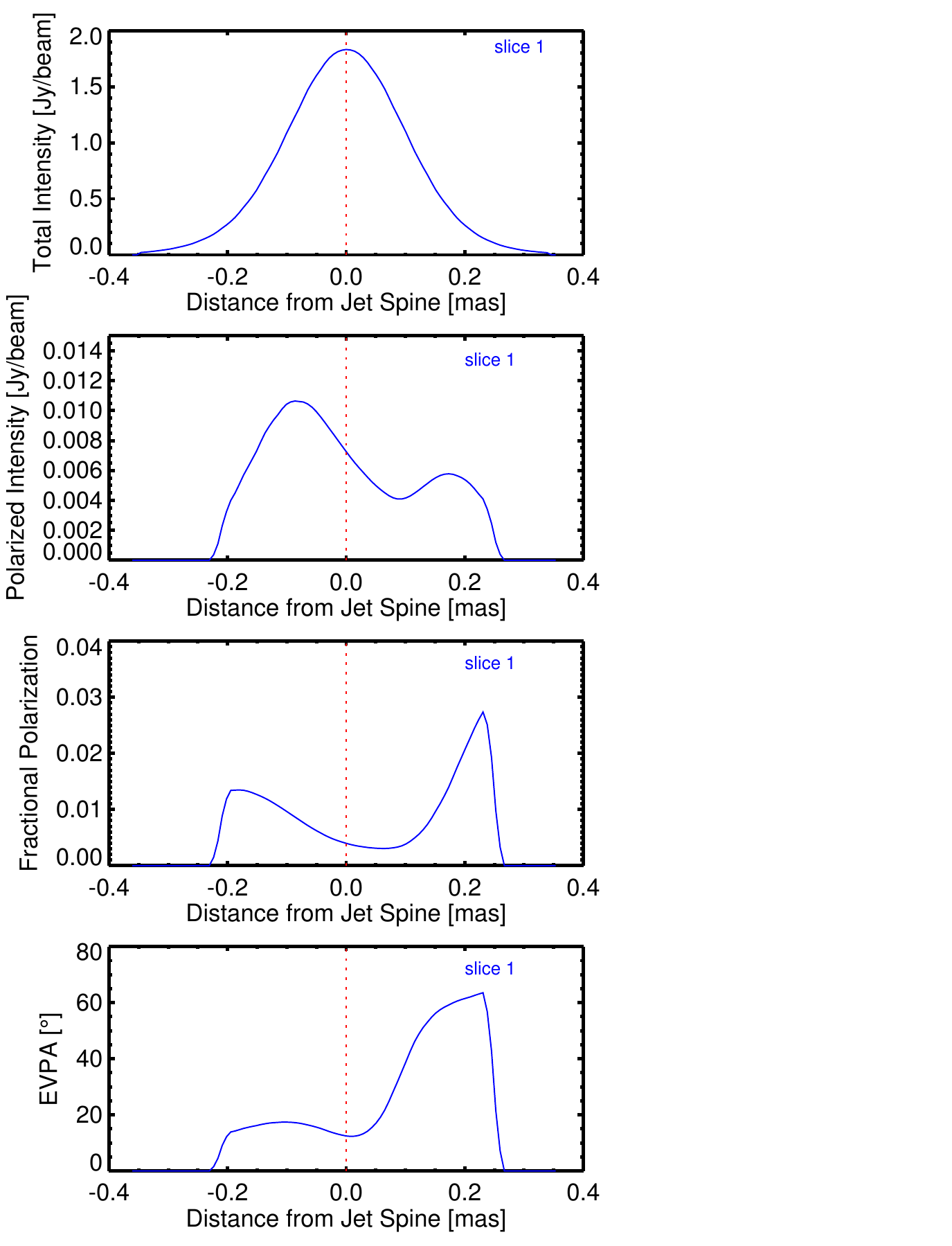}}
  \end{center}
  \caption{\label{fig7}Variations of the emission parameters transverse to the jet axis (dashed red line) for slice 1 (shown in Figure 6): total intensity (top panel), polarized intensity (upper middle panel), fractional polarization (lower middle panel), and EVPA (bottom panel).  The fractional polarization increasing toward the edges of the jet is a predicted polarimetric signature of a jet sheath (as discussed in \S5), with the EVPA transverse to the axis.}
\end{figure}

A complication in detecting the polarimetric signature of a jet sheath within a blazar is discussed by \citet{clausen-brown13}.  Due to the effects of Doppler beaming, the relativistic spine of the jet ($\Gamma_{\rm{spine}} \sim 20$) is intrinsically far brighter than the surrounding sheath (which is usually assumed to be non- to mildly-relativistic: $\Gamma_{\rm{sheath}} \sim 2$).  To overcome this vast difference in intrinsic brightness, several authors have implemented a method of ``stacking'' radio maps (see, e.g., \citealt{fromm13}; \citealt{zamaninasab13}).  This stacking method involves aligning individual radio images of a blazar based on the location of the radio core (which is assumed to be an essentially stationary feature within the jet at a given frequency).  The images are then added together, with the expectation that noise and transient phenomena (such as superluminal blobs) will smooth out in the stacked map, and, in contrast, any standing features within the jet (such as a jet sheath) will be highlighted as more images are added to the stacked map.  With this in mind, we stacked VLBA total intensity and polarization images of PKS 1510$-$089 at 43 GHz in order to see whether the polarimetric signature of a jet sheath is present in this blazar.  In particular, we selected images from twenty epochs of observation from 2008 to 2013 made with VLBA data from the VLBA-BU-BLAZAR program (see www.bu.edu/blazars/VLBAproject.html and \citet{jorstad13} for a general description of the observations and data analysis).  The chosen epochs correspond to periods of relatively quiescent jet activity.  Our stacked map for PKS 1510$-$089 is shown in Figure \ref{fig6}.  We do see an increase in the degree of polarization toward the edges of the jet.  The EVPAs plotted in Figure \ref{fig6} are inclined to the jet spine axis (highlighted in red) and are predominantly perpendicular to that axis (as predicted by the sheath scenario) in both the southern half and the upper northern half of the jet.  We examine ten transverse slices through the data (shown in blue).  The variations of the Stokes parameters along the first slice are shown in Figure \ref{fig7}.  We indeed find that the fractional polarization increases toward the edges of the jet, as predicted by the sheath scenario.  These two features, namely, the increase in fractional polarization and the orientation of that polarization perpendicular to the jet axis, are consistent with a sheath in PKS 1510$-$089.  The variations of the Stokes parameters along the other nine slices are qualitatively similar to the first, with the exception that after the fifth slice we no longer see signs of a sheath component in the southern half of the jet. 

\begin{figure}
\epsscale{1.15}
\plotone{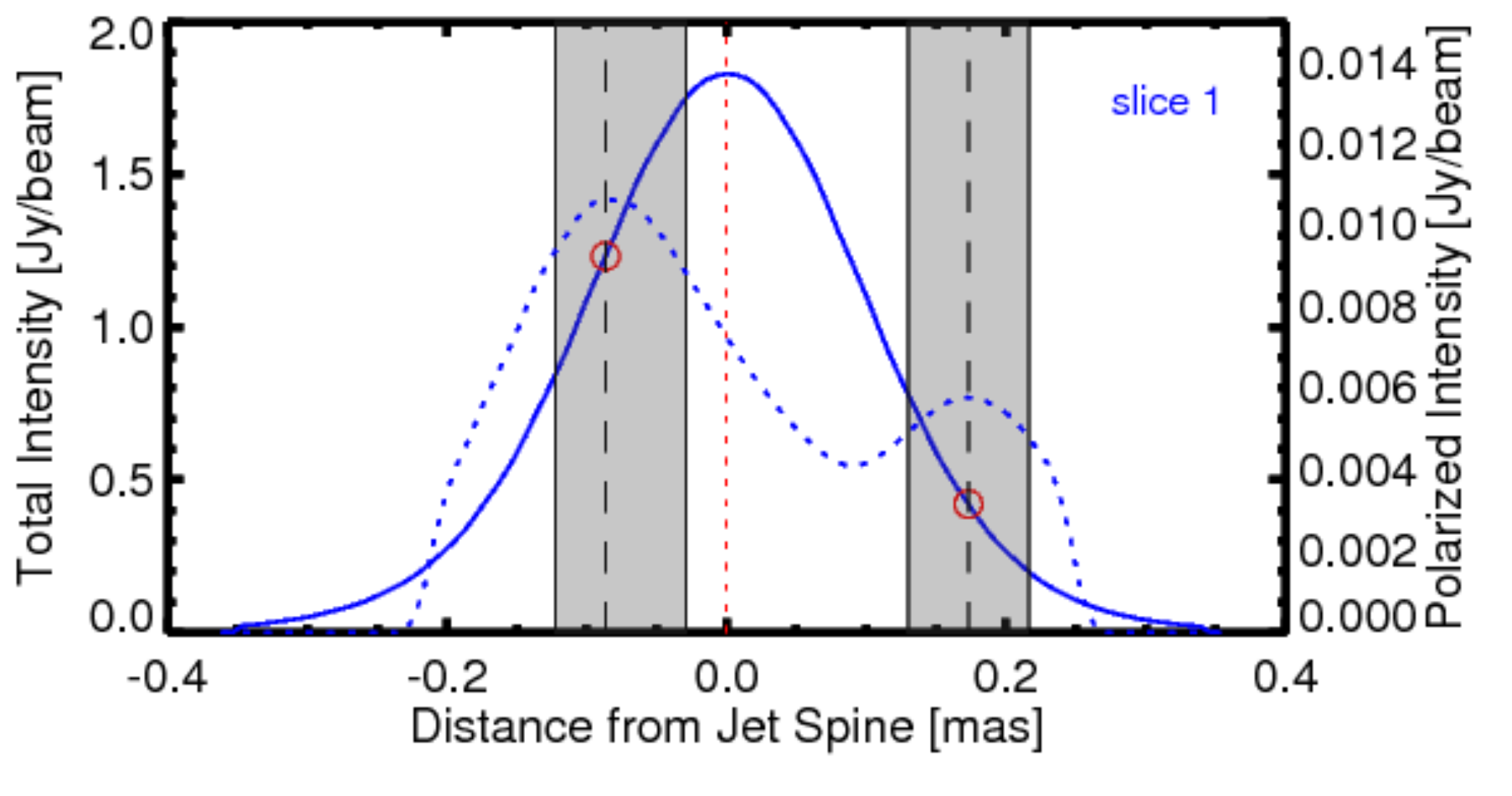}
\caption{\label{fig8}Illustration of the procedure used to determine both the location of the sheath within the jet and its contribution to the observed emission by utilizing the total and polarized intensity profiles (shown in Figure \ref{fig7}) of the slices taken through our stacked map (shown in Figure \ref{fig6}).  The polarized intensity profile along slice 1 (dashed blue line corresponding to the right-hand axis) is overlaid upon the total intensity profile for slice 1 (solid blue line corresponding to the left-hand axis).  We use this double-peaked profile to highlight the likely location of the sheath within each slice (shaded gray regions).  The sheath's contribution to the jet's total intensity profile is estimated by the values of total intensity (red circles) that are cospatial with the peaks of the polarized intensity profile (demarcated by dashed vertical lines).  This procedure is then repeated for each slice, thus tracing out the sheath (shaded gray regions) shown in Figure \ref{fig6}.}
\end{figure}

The total and polarized intensity profiles along the slices (highlighted in Figure \ref{fig7}) are used to determine the nominal location of the sheath within the jet, as illustrated in Figure \ref{fig8}.  In particular, we align the total intensity and polarized intensity profiles for each slice and use the double-peaked nature of the polarized intensity (a signature of the jet sheath) to highlight the location of the sheath within each slice (the shaded gray regions shown in Figure \ref{fig8}).  The width of the sheath on each side of the slice is determined, arbitrarily, by the location where the polarized flux falls below 0.85 times the peak value (the solid black vertical lines in Figure \ref{fig8}).  This percentage yields a conservative estimate of the extent of the sheath.  The total intensity of the sheath plasma on each side of the jet is approximated by the values of the total intensity profile (the red circles in Figure \ref{fig8}) that are aligned with the peaks of the polarized intensity profile (the dashed black vertical lines in Figure \ref{fig8}).  We then repeat this procedure for each slice throughout our stacked map, thus tracing out the sheath (shaded gray regions shown in Figure \ref{fig6}).  

\begin{figure*}
  \setlength{\abovecaptionskip}{-6pt}
  \begin{center}
    \scalebox{1.0}{\includegraphics[width=2.0\columnwidth,clip]{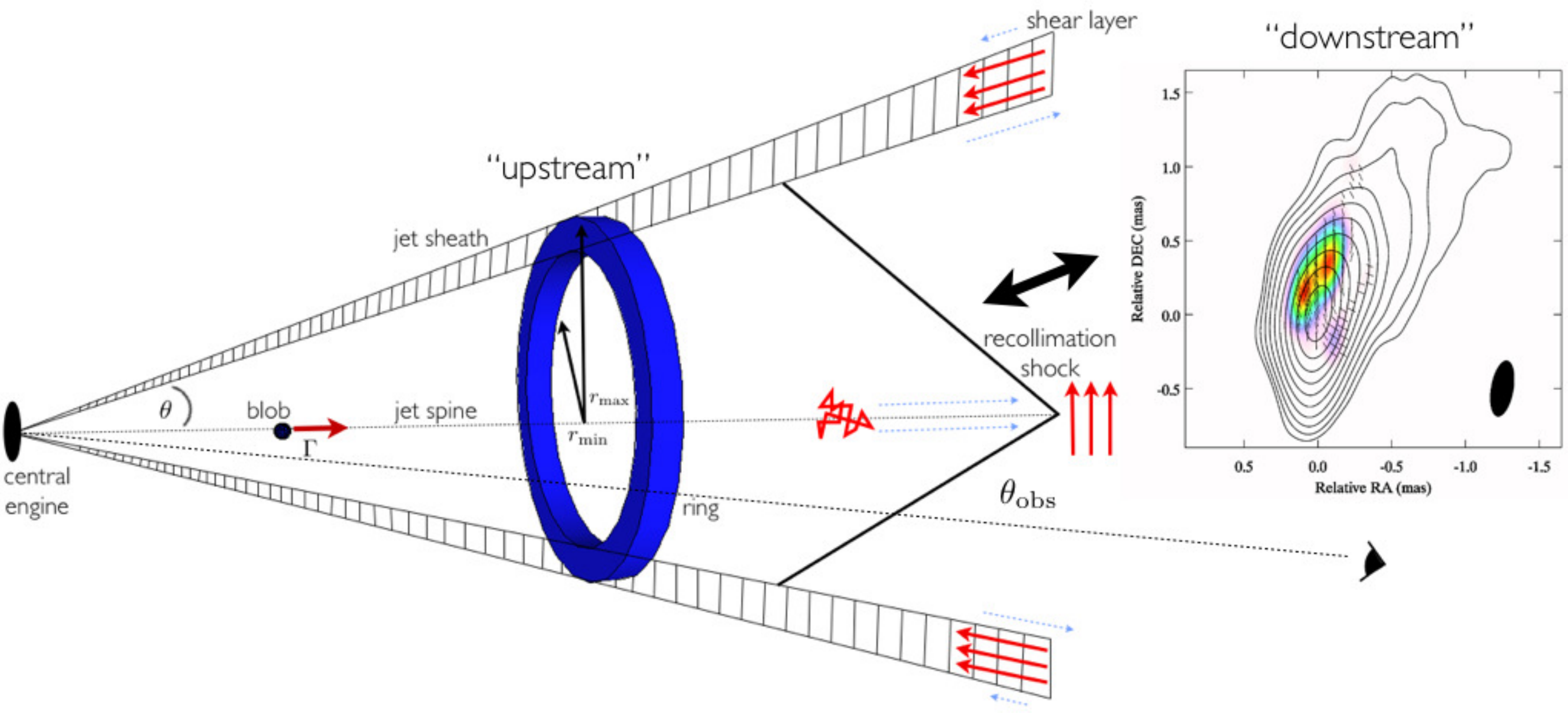}}
  \end{center}
  \caption{\label{fig9}A schematic of the relative locations along the jet of both the ring of shocked sheath plasma in our model and the location of the radio core/sheath detected further downstream in the stacked radio images of PKS 1510$-$089 shown in Figure \ref{fig6}.  As discussed in \S3, we posit that the ring is located $\sim 4 ~ \rm{pc}$ from the central engine while the radio core in the stacked map is located further downstream at a scale of $\gtrsim 10 ~ \rm{pc}$ from the central engine.  We propose that the radio core in PKS 1510$-$089 is associated with a recollimation shock that compresses initially tangled magnetic field along the spine of the jet (shown in red) an orders that field perpendicular to the jet axis (the red vectors just to the right of the recollimation shock).  In contrast to the spine, velocity shear between the sheath and the ambient medium (blue vectors denote relative speed) aligns the magnetic field lines on the outer edges of the jet to be roughly parallel to the jet axis resulting in the spine-sheath polarization signature we detect in our stacked map of PKS 1510$-$089 (shown to the right).}
\end{figure*}

An estimate of the bolometric luminosity ($L_{\rm bol}$) of the sheath in the observer's frame is computed by adding all of the flux contained within the gray shaded regions of our stacked map (Figure \ref{fig6}).  Based on the distance to PKS 1510$-$089, we convert this inferred sheath flux at 43 GHz into a spectral luminosity ($L_{\nu}$) and assume a power-law of the form $L_{\nu} = C \nu^{-\alpha}$.  After solving for the constant of proportionality ($C$), we then integrate to obtain:
\begin{equation} \label{eq17}
L_{\rm bol} = \int_{\nu_{\rm{min}}}^{\nu_{\rm{max}}} L_{\nu} ~ \! \mathrm{d} \nu \sim 3 \times 10^{45} ~ \rm erg ~ \rm s^{-1},
\end{equation}
where we have assumed a spectral index of $\alpha \sim 1.0$ (in rough agreement with the mean spectral index of a large sample of flat spectrum radio quasar jets found by \citealt{hovatta14}) and limits of integration of $\nu_{\rm min} = 10^{9} ~ \rm Hz$ and $\nu_{\rm max} = 5 \times 10^{13} ~ \rm Hz$.  It is interesting to note the similarity between this value and the value obtained in \S3 for the upstream sheath luminosity using Equation (\ref{eq16}).  The above estimate indicates that the sheath within PKS 1510$-$089 is potentially a very important source of seed photons even at parsec scales along the jet.  As a further comparison, we compute the bolometric luminosity (in the observer's frame) of the ring of sheath plasma in our model assuming, in contrast to the simulation, that the ring moves along the jet with a bulk Lorentz factor of $\Gamma_{\rm ring} = 2$.  We find $L_{\rm bol}^{\rm ring} \sim  3 \times 10^{44} ~ \rm erg ~ \rm s^{-1}$, roughly an order of magnitude smaller than the above sheath estimate. 

Finally, a crude estimate of both the inner ($r_{\rm min}$) and outer ($r_{\rm max}$) radii of the sheath was made by averaging the values obtained for these parameters relative to the jet spine (demarcated by the dashed red vertical line in Figure \ref{fig8}) for each slice through the stacked map.  We find that $r_{\rm min} \sim 0.4 ~ \rm pc$ and $r_{\rm max} \sim 0.8 ~ \rm pc$.  These values are roughly a factor of $4.5$ larger than the inner and outer radii used in our model of the ring (see Table \ref{tab1}) that we posit to exist upstream of the region highlighted in Figure \ref{fig6}.  It is interesting to note that the opening angle of the sheath used in our model ($\theta_{\rm sh} \equiv \theta_{\rm max} - \theta_{\rm min} \sim 1.29^{\circ} ; \theta_{\rm min} = 1.29^{\circ} ~ \rm and ~ \theta_{\rm max} = 2.58^{\circ}$ as discussed in \S3) predicts that $r_{\rm min} = z ~ \rm tan( \theta_{\rm min} ) \sim 0.4 ~ \rm pc$ and $r_{\rm max} = z ~ \rm tan( \theta_{\rm max} ) \sim 0.8 ~ \rm pc$ for the sheath at a distance of $z = 17 ~ \rm pc$ from the central engine, which is roughly the scale where the radio core is situated in our stacked map at 43 GHz \citep{marscher10}.

Although the plasma in this sheath needs to move relativistically for beaming to allow us to detect it, given this transverse velocity gradient, we infer the presence of an even slower sheath upstream, where the ring in our model could be located (as discussed in \S3 and illustrated in Figure \ref{fig9}).  An enhancement, such as a ring of shocked plasma, in this more slowly moving sheath of plasma could potentially supply the seed photons that our model requires to explain the observed orphan flares within PKS 1510$-$089.

\section{Summary and Conclusion}

We have developed a model to explain the origins of orphan $\gamma$-ray flares from blazars and have applied this model to a specific orphan flare within the blazar PKS 1510$-$089.  In this model, a shocked segment of a slow jet sheath provides a localized source of seed photons that are inverse-Compton scattered by electrons contained within a blob of plasma moving relativistically along the spine of the jet.  This inverse-Compton scattering results in the production of a sharply peaked orphan  $\gamma$-ray flare.  Synthetic light curves produced by this model are able to reproduce the general features of the observed $\gamma$-ray light curve during an orphan flare event associated with the ejection of a superluminal knot from the base of the jet of PKS 1510$-$089.  The model is also able to reproduce the double-peaked profile seen in the SEDs of most blazars.  This model can potentially be used to discern the nature of the motion of the blob along the spine of the jet based on the shape of the observed orphan flare.  We have stacked radio images of PKS 1510$-$089 obtained with the VLBA at 43 GHz at multiple epochs.  We find the polarimetric signature of a sheath in our stacked map, thus lending observational support to the plausibility of this model of orphan $\gamma$-ray flares within blazars.  In the future, we plan to extend our modeling efforts in an attempt to reproduce orphan $\gamma$-ray flares observed in the blazars 3C 273, 3C 279, and 4C 71.07.  We also plan to create stacked maps of the jets of these objects to investigate whether we can detect additional spine-sheath polarimetric signatures.
 
\subsection*{Acknowledgements}

Funding for this research was provided by a Canadian NSERC PGS D2 Doctoral Fellowship and by NASA Fermi Guest Investigator grants NNX11AQ03G, NNX12AO79G, NNX12AO59G and NNX14AQ58G.  The authors are grateful to Karen Williamson for supplying the multi-wavelength data used in Figure \ref{fig3} and to the anonymous referee for a thorough and helpful review.  The VLBA is an instrument of the National Radio Astronomy Observatory.  The National Radio Astronomy Observatory is a facility of the National Science Foundation operated under cooperative agreement by Associated Universities, Inc.

\end{document}